\documentclass[conference]{IEEEtran}
\IEEEoverridecommandlockouts
\usepackage{cite}
\usepackage{amsmath,amssymb,amsfonts}
\usepackage{algorithmic}
\usepackage{graphicx}
\usepackage{textcomp}
\usepackage[dvipsnames]{xcolor}
\def\BibTeX{{\rm B\kern-.05em{\sc i\kern-.025em b}\kern-.08em
    T\kern-.1667em\lower.7ex\hbox{E}\kern-.125emX}}

\usepackage[utf8]{inputenc}
\usepackage[T1]{fontenc}
\usepackage{tikz}
    \usetikzlibrary{calc,matrix, positioning}
    \usetikzlibrary{shapes.symbols}
    \usetikzlibrary{decorations.pathreplacing,calligraphy}
    \usetikzlibrary{arrows.meta}
    \usetikzlibrary{shapes.geometric}
\usepackage{fontawesome5}
\usepackage{multicol}
\usepackage{multirow}
\usepackage{booktabs}
\usepackage{hyperref}
\usepackage{csquotes}
\usepackage{subfigure,bm}
\usepackage{eurosym,nicefrac}
\usepackage{microtype}

\newcommand{\encom}[1]{``#1''}
    
\begin{document}

\title{Large-Scale (Semi-)Automated Security Assessment of Consumer IoT Devices -- A Roadmap\\
\thanks{The financial support by the Austrian Federal Ministry of Labour and Economy, the National Foundation for Research, Technology and Development and the Christian Doppler Research Association is gratefully acknowledged.}
}

\author{
\IEEEauthorblockN{Pascal Schöttle~~Matthias Janetschek~~Florian Merkle~~Martin Nocker~~Christoph Egger}
\IEEEauthorblockA{\textit{Josef Ressel Centre for Security Analysis of IoT Devices} \\
Digital Business \& Software Engineering\\
MCI -- The Entrepreneurial School\\
Innsbruck, Austria \\
\{pascal.schoettle~|~matthias.janetschek~|~florian.merkle~|~martin.nocker~|~christoph.egger\}@mci.edu}
}

\maketitle

\begin{abstract}
The Internet of Things (IoT) has rapidly expanded across various sectors, with consumer IoT devices - such as smart thermostats and security cameras - experiencing growth. Although these devices improve efficiency and promise additional comfort, they also introduce new security challenges. Common and easy-to-explore vulnerabilities make IoT devices prime targets for malicious actors.

Upcoming mandatory security certifications offer a promising way to mitigate these risks by enforcing best practices and providing transparency. Regulatory bodies are developing IoT security frameworks, but a universal standard for large-scale systematic security assessment is lacking. Existing manual testing approaches are expensive, limiting their efficacy in the diverse and rapidly evolving IoT domain.

This paper reviews current IoT security challenges and assessment efforts, identifies gaps, and proposes a roadmap for scalable, automated security assessment, leveraging a model-based testing approach and machine learning techniques to strengthen consumer IoT security.
\end{abstract}


\section{Introduction}

The Internet of Things (IoT) has become an integral part of modern society, with applications in various sectors such as smart manufacturing~\cite{yang2019internet}, transport and logistics~\cite{song2020applications}, agriculture~\cite{farooq2020role}, or healthcare~\cite{kashani2021systematic}. 
In addition to these commercial use cases, consumer IoT (CIoT) devices, such as smart thermostats~\cite{ozgur2018iot}, security cameras~\cite{abdalla2020testing}, and connected home appliances~\cite{shashank2021power}, are experiencing particularly rapid growth. A study published by Data Bridge Market Research estimates a global Compound Average Growth Rate for CIoT of 17.43\% in the forecast period of 2021 to 2028~\cite{DataBridgeMarketResearch.2021}. 
Connectivity of these devices facilitates automation, convenience, and efficiency, but also introduces new attack surfaces and security challenges~\cite{schiller2022landscape}.

Compared to traditional IT systems, IoT devices face distinct security challenges:
Generally, IoT devices are resource-constrained, limiting the feasibility of implementing robust security mechanisms~\cite{vcolakovic2018internet}. Additionally, the Open Web Application Security
Project (OWASP) IoT Project has identified several critical issues 
in their \textit{Top 10} list, making these devices vulnerable to unauthorized access~\cite{owasptopten}. 
Their integration with critical infrastructure~\cite{9165889} and the handling of personal data~\cite{torre2016framework} further make IoT devices an attractive attack target for malicious actors. 
High-profile incidents, such as the Mirai botnet and several breaches in industrial control systems~\cite{antonakakis2017understanding} have displayed the high risk associated with insecure IoT devices.
Automated security assessment and certification could address these security implications by providing transparency to customers and forcing manufacturers to apply good security practices. Accordingly, governments and regulatory organizations, including the EU, European Union Agency for Cybersecurity (ENISA), European Telecommunications Standards Institute (ETSI), and US National Institute of Standards and Technology (NIST), are introducing IoT security frameworks to mitigate emerging risks. 
However, no security assessment standard currently exists that systematically evaluates IoT security at scale. Several challenges arise in defining such a standard, including the scope of testing -- whether it should cover hardware, software, or the entire system. The responsibility for security assessment is also unclear, as it could fall to manufacturers, independent IT security labs, or regulatory bodies. Additionally, the highly heterogeneous IoT landscape introduces an additional challenge regarding the scalability of security assessment efforts.
Manual security assessment processes are slow, costly, and impractical given the scale and diversity of IoT devices. To address this, the industry requires semi- or fully automated security evaluation techniques that enable scalable, repeatable, and objective assessments. Initial efforts in this direction have already been made~\cite{kaksonen2024automating}. Further, machine learning has been applied to software testing~\cite{ajorloo2024systematic} and web application security testing~\cite{appelt2018machine}, and may contribute to increasing security test automation.

In this work, we review the current state of the art on IoT security challenges and security assessment efforts. We identify research gaps between security requirements, real-world vulnerabilities, and existing testing initiatives. 
By doing so, we are able to propose a roadmap outlining future research directions to address the key challenges of scalable IoT security assessment for consumer IoT devices.
The remainder of this work is structured as follows: The subsequent sections~\ref{sec:iot} and~\ref{sec:iotsec} introduce the relevant background on the IoT in general and its security properties specifically. In Section~\ref{sec:roadmap}, we present open challenges and our proposed roadmap before we conclude this work in Section~\ref{sec:conclusion}

\section{The Internet of Things}
\label{sec:iot}

A key goal of digitalization is to link real-world objects, also called things, with the digital world of the Internet. 
These things transfer data to the Internet, which usually represent physical or chemical variables measured by sensors. This eliminates the strict separation between the real world and virtual worlds. Physical objects can have a virtual equivalent, which offers a wide range of new possibilities but also new threats. 

In the scientific literature, several IoT definitions appear, capturing different viewpoints. A rather general definition from Lee and Lee~\cite{Lee.2015}, describes IoT as \encom{a new technology paradigm envisioned as a global network of machines and devices capable of interacting with each other}.

Dorsemain et al.~\cite{Dorsemaine.2015} define the IoT as \encom{a group of infrastructures, interconnecting connected objects and allowing their management, data mining and the access to data they generate} where the connected objects are \encom{sensor(s) and/or actuator(s) carrying out a specific function that are able to communicate with other equipment}.

All definitions include the connection of devices that collect environmental data and transfer them to other devices or central IT systems. This environmental data is usually collected via sensors and includes, for example, humidity, temperature, pressure, CO2 emission, and proximity. Other widespread sensors include gyroscopes, accelerometers, and location (e.g., GPS coordinates). 
The data can further be processed, remotely or directly on the device, and decisions can be derived, which are executed by actuators.

\subsection{Industrial and Consumer Internet of Things}

IoT can be divided into several streams, primarily industrial IoT (IIoT) and consumer IoT (CIoT). The key distinction lies in their application contexts.
IIoT builds upon the concept of the Industrial Internet, first introduced by General Electric, emphasizing the interconnection of industrial machines, sensors, and actuators within broader industrial networks. It is particularly relevant to industrial production~\cite{Sadeghi.2015}.
CIoT, in contrast, focuses on consumer products, ranging from simple devices to sophisticated smart home systems~\cite{EuropeanCommission.2021}. Examples include smart bulbs and robot vacuums. Consumers typically choose from various comparable products, which often rely on WiFi connectivity. A home WiFi setup usually involves a router provided by an Internet service provider. In most cases, consumers cannot select the router model, and routers themselves qualify as IoT devices.
CIoT encompasses numerous formerly standalone devices now integrated with network connectivity, such as smart locks, smart watches, robot vacuums, washing machines, smart meters, (semi-) autonomous cars, and more. Additionally, smartphones and tablets, equipped with various sensors, fall within the broader IoT category.

Despite differing use contexts, IIoT and CIoT share fundamental properties, and some devices, such as smart locks or bulbs, are used in both domains. IoT devices, consisting of physical components and controllers with sensors and actuators, form the foundation of any IoT system. Compared to non-IoT devices, they often have limitations in computational power, memory, power supply, and data rates~\cite{Alladi.2020}. IoT architectures are designed to address these constraints.

In the area of home security and smart home solutions, CIoT enhances automation and energy efficiency while also offering security features such as alerts and residential monitoring~\cite{Hoque.2019}. Healthcare and ambient assisted living applications primarily support elderly or sick individuals in daily routines and enable emergency detection, such as fall detection, as well as baby monitoring with real-time updates for parents~\cite{Marques.2021}. Another key domain is personal fitness and sports, where wearables and smart sports equipment provide users with performance data and insights for training optimization~\cite{Passos.2021}. Additionally, personal asset tracking has become an increasingly relevant application, allowing individuals to locate lost or misplaced belongings such as mobile phones, laptops, keys, or passports~\cite{Alsubhy.2020}.

Furthermore, IoT also serves as a foundational technology for Smart Cities, where public service providers integrate IoT-based infrastructures into urban environments. Unlike CIoT, Smart City systems operate at the municipal level, often affecting individuals indirectly, even if they do not own IoT devices themselves~\cite{Gupta.2019}. 
These initiatives aim to improve resource efficiency, transportation, waste management, and utility monitoring through interconnected sensor networks and data-driven decision-making~\cite{Harrison.2010}. The integration of smart water, electricity, and traffic management systems exemplifies the role of IoT in optimizing urban services~\cite{Elmustafa2019InternetOT}. However, the complexity of Smart Cities increases with the heterogeneity of connected systems, particularly as smart homes become part of larger urban infrastructures~\cite{Montori.2018}.

Security and privacy are critical concerns across all IoT applications~\cite{Alladi.2020,Johnson.2020,Xenofontos.2021}. 
Consumers often rely on pre-implemented security features and lack the ability to modify them. Increasing awareness of security and privacy risks remains a key challenge in strengthening the IoT ecosystem.

\subsection{IoT Architectures}
\label{sec:architectures}
As mentioned above, IoT is not only about smart things (physical objects with sensors and actuators). The term IoT includes the data transfer from things to applications or services that make use of these data.  Since IoT offers a wide range of applications and services, IoT solutions architectures support system design and development.

\begin{figure*}[ht]
    \centering
    \includegraphics[width=.6\textwidth]{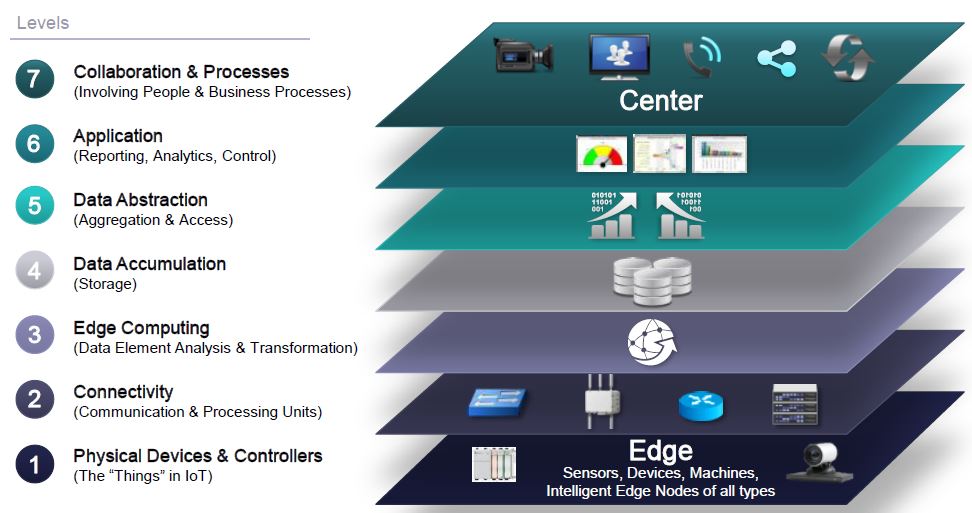}
    \caption{IoT World Forum Reference Model, based on~\cite{ElHakim.2018}}
    \label{fig:IoT_World_Forum_Reference_Model}
\end{figure*}

Several IoT architectures and related security concepts have been developed in the last years, e.g.~\cite{Singh.2020}.
They all have four major building blocks or layers in common:
1) the things which comprise sensors and actuators; 2) connectivity, network gateways, and internet connections, sometimes with integrated data pre-processing capabilities; 3) the middleware, such as IoT platforms or data processing at the edge of the network; 4) the applications which are provided in on-premise data centers or the cloud.

At the IoT World Forum 2014, Cisco, IBM, and Intel have presented an IoT solutions reference model~\cite{ElHakim.2018}. This reference model is more fine granular and consists of seven layers, from the \encom{things layer} to the \encom{business layer}, as depicted in Figure~\ref{fig:IoT_World_Forum_Reference_Model}.

\subsection{Data Transmission Technologies}
\label{sec:datatransmission}

IoT devices predominantly rely on wireless data transmission for flexible placement and mobility. This section outlines key wireless protocols commonly used in IoT applications.

IEEE 802.11 (Wi-Fi) defines PHY and MAC protocols for WLANs, operating in the 2.4 GHz and 5 GHz bands, with Wi-Fi 6 extending to 6 GHz. Data rates range from a few Mbps to 9608 Mbps, with a typical range of 100 meters. 
Substandards such as 802.11af, 802.11ah, and 802.11ba enhance Wi-Fi's applicability for IoT by improving range and energy efficiency. 

Bluetooth operates in the 2.4 GHz band with data rates up to 3 Mbps and ranges up to 100 meters, though specialized hardware extends this~\cite{BluetoothRange}. Bluetooth Low Energy (BLE) optimizes energy consumption and is widely used in IoT.
Bluetooth Mesh enables multi-hop networking, improving coverage for IoT applications.

LoRaWAN, a prominent LPWAN protocol, operates in sub-gigahertz bands such as 868 MHz in Europe, with data rates between 250 bits/s and 50 kbits/s. It supports transmission distances up to 2 km in urban areas and 40 km in rural settings, making it ideal for low-power, long-range IoT applications such as smart city infrastructure. 

IEEE 802.15.4 underpins low-rate WPANs, forming the basis for Zigbee and Thread. Zigbee enables multi-hop mesh networking with low power consumption but suffers from vendor-specific interoperability limitations. Thread, an IPv6-based protocol, enhances scalability and security. 

Z-Wave, operating in sub-gigahertz bands, provides data rates up to 100 kbits/s with a range of 100 meters. It emphasizes interoperability and, though less common than Zigbee, remains relevant in consumer IoT. 

Matter, a new application-layer protocol maintained by the Connectivity Standards Alliance, aims to standardize IoT interoperability by leveraging Bluetooth LE for setup and Wi-Fi or Thread for operation. To counter vendor fragmentation, it requires device vendors to certify their devices.

Cellular networks (2G-5G) support IoT applications requiring mobility. While 2G (up to 384 kbit/s) is being phased out, 4G remains the dominant choice. 5G offers superior speed (up to 20 Gbit/s) and low latency but faces adoption challenges due to cost and power consumption.

\section{IoT Security}
\label{sec:iotsec}

Given the prevalence of IoT devices and their influence on our daily lives now and the likely development thereof in the near future~\cite{acmtechbriefsc}, it is understandable that more and more users, researchers, companies, and governments express security concerns about IoT, e.g.,~\cite{siboni2019security}. One example is the European Commission's delegated act to the Radio Equipment directive\footnote{The complete delegated act is available at: \url{https://ec.europa.eu/growth/document/download/492e4668-f9c2-495c-ac11-4379dd2533d9_en}}. This act states: \encom{As mobile phones, smart watches, fitness trackers and wireless toys are more and more present in our everyday life, cyber threats pose a growing risk for every consumer} and it aims: \encom{to make sure that all wireless devices are safe before being sold on the EU market. This act lays down new legal requirements for cybersecurity safeguards, which manufacturers will have to take into account in the design and production of the concerned products.} How to control compliance with these regulations is neither openly addressed nor solved.
Another very recent example is the European Commission's \textit{Cyber Resilience Act} (CRA)~\cite{EU-CRA}, which was finally accepted in October 2024 and will come fully into force in December 2027.

Although the basic protection goals from information security, namely confidentiality, integrity, and availability, are easily transformable to the domain of IoT devices, unfortunately, the methods to achieve them are not. Furthermore, in IoT, security is strongly connected to safety. Whether accidental or malicious, interference with the controls of medical devices, such as a pacemaker, a self-driving car, or a nuclear reactor poses a serious threat to human life.

Even more, the sheer abundance of IoT devices connected to the Internet, paired with their oftentimes weak security measures, makes these devices worthwhile targets for cybercriminals as they can be used in, for example, distributed denial-of-service (DDoS) attacks. Probably the most popular example of such an attack using mainly IoT devices is the DDoS attack on, among others, the website of computer security journalist Brian Krebs in September 2016 using the so-called \encom{Mirai botnet}~\cite{antonakakis2017understanding}. This botnet consisted of mainly IP cameras, printers, routers, and other IoT devices. Many similar examples since then highlight the need for more secure IoT devices that must not be easily taken over by adversaries.

\subsection{Dimensions of IoT Security}

\paragraph{Major IoT Security Challenges}
The various deployment areas of IoT, from wearable devices to cyber-physical systems (CPS) in Industrial IoT, create a highly heterogeneous environment. This heterogeneity spans sensor types, transmission protocols, middleware, and applications. However, a common constraint across IoT devices is limited computational power, memory, and energy supply. As a result, traditional security measures, such as resource-intensive authentication and encryption, cannot be directly applied to IoT systems.

\paragraph{Access control and device authentication}
IoT devices are typically designed for low power consumption, with constrained processors and limited connectivity. Many are \encom{headless}, meaning no human input is available for authentication. Implementing access control and authentication is challenging due to the lack of processing power, making standard encryption techniques infeasible. Solutions must balance security, usability, and hardware limitations without introducing weak or easily breakable parameters.

\paragraph{Lifecycle, future-proofing, updates}
Many companies neglect long-term support for IoT devices, leading to millions of unpatched, insecure devices. If even major smartphone vendors discontinue updates after a few years, the situation is likely worse for cheap IoT devices, which may remain connected to the Internet for extended periods. Additionally, household appliances such as refrigerators and washing machines, now increasingly IoT-enabled, typically last 20+ years—far longer than the expected support window for software updates. Manufacturers have little financial incentive to provide updates for old devices, leaving IoT security dependent on either inherently secure designs or continuous updates, both of which are unrealistic. Bruce Schneier highlights that \encom{neither the buyer nor the seller cares}\footnote{\url{https://www.schneier.com/essays/archives/2016/10/we_need_to_save_the_.html}}, 
making IoT insecurity a negative externality akin to invisible pollution. The same issue applies to IIoT and Smart Cities, where infrastructure components are expected to remain in use for decades.

\paragraph{Physical Accessibility}
From production to disposal, IoT devices may be subject to tampering. Traditional IT security assumes adversaries lack physical access, but IoT devices are frequently deployed in remote, unattended, or public locations, making them highly accessible to attackers. Once physical access is granted, a range of attack vectors, including side-channel attacks, becomes possible. Contactless techniques, such as monitoring electromagnetic emissions or power fluctuations, can also be used to extract sensitive data. Smart Cities and businesses using CIoT face additional risks, as an adversary with physical access to an IoT device may gain entry to internal networks and sensitive data.

\paragraph{Data Transmission}
IoT devices commonly use wireless communication networks for their flexibility, but wireless networks are inherently insecure due to their shared nature. Two key attack vectors arise from this. One is provisioning attacks, where many IoT devices lack interfaces for human authentication and instead rely on automated provisioning schemes. Zigbee, for example, permits device onboarding through a simplified process requiring minimal user interaction. \cite{CheckPoint2020} describes an attack exploiting this scheme, tricking users into initiating a join process that allows a malicious device to infiltrate the network. Another major threat is wireless transmission interception, where attackers with suitable hardware can intercept or inject packets into IoT networks. 

Finally, IoT security is further complicated by the complexity of provisioning and encryption schemes. While they may appear simple externally, they involve intricate protocols with numerous configurable parameters. This complexity increases the likelihood of implementation errors and misconfigurations, which create security vulnerabilities.

\subsection{State of the Art}
\label{sec:stateoftheart}

In recent years, scientific research on IoT security has increased significantly~\cite{Xenofontos.2021,Neshenko2019,mahmoud2015internet}. 
Various organizations, both governmental and non-governmental, have also published security recommendations and lists of common vulnerabilities. Notable examples include the OWASP Top 10 IoT security vulnerabilities~\cite{owasptopten}, the ENISA baseline security recommendations for IoT~\cite{ENISA2017baseline}, the ETSI standard for cybersecurity in IoT, which establishes a security baseline for consumer IoT products~\cite{ETSI2021}, and the EU's CRA~\cite{EU-CRA}.

These recommendations likely reflect real-world security challenges better than some research efforts that, e.g., focus on lightweight post-quantum encryption for IoT devices~\cite{Bouguettaya2021}. In practice, a large portion of deployed IoT devices suffer from weak, guessable, or hardcoded passwords (No. 1 on OWASP’s list) or lack secure update mechanisms (No. 4), while few are realistically threatened by quantum-enabled attacks.

With increasing attention to IoT security, proposals for security by design are gaining traction. These include approaches such as computer-aided design~\cite{xu2014security}, system hardening and monitoring~\cite{choi2018system}, and case study-based security recommendations~\cite{Alladi.2020}. However, implementation remains limited due to weak consumer demand for secure devices, reducing manufacturers' incentives to adopt these measures. As recent research suggests that consumer awareness is rising and comes together with an increased willingness to actually pay more for more secure IoT devices~\cite{emami2023consumers}, this also might increase manufacturers' interest in security by design methods.

Machine learning (ML) is emerging as a promising tool for assessing IoT security. ML-based methods are being integrated into security testbeds~\cite{siboni2019security} and are applied to attack detection and prediction~\cite{al2020survey}. Additionally, ML and artificial intelligence (AI) are seen as essential for protecting security and privacy in Smart Cities~\cite{acmtechbriefsc}.

\section{The Roadmap -- Challenges and Open Research Questions for Automated IoT Security Assessment}
\label{sec:roadmap}

Besides the official IoT security recommendations by well-known organizations such as the European Commission, ENISA, and ETSI, there is still no reliable and reputable security assessment or certification authority covering a broad range of IoT devices.
As a result, consumers and companies cannot rely on IoT devices to at least meet baseline security recommendations~\cite{ENISA2017baseline,ETSI2021}. This absence is not due to a lack of proposals for testing techniques~\cite{Medhat2019,voas2018testing}, security testbeds~\cite{siboni2019security}, or even certification frameworks~\cite{Matheu2019} for IoT. 
Instead, it may be linked to fundamental questions such as what exactly constitutes IoT, who is responsible for security assessment and certification, and what exactly should be certified -- a simple sensor, an entire device, or the device along with a specific firmware version.
These and similar issues have been referred to as \encom{IoT's Certification Quagmire}~\cite{Voas2018} and remain unresolved so far.

\subsection{Security Assessment Process}

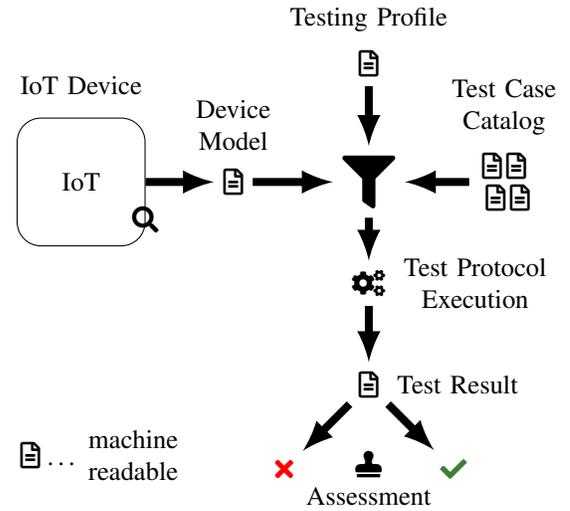
\begin{figure}[htb]
    \centering
    \begin{tikzpicture}[
                >=latex,
                myarrow/.style={->,>=latex,line width=3pt},
                scale=.45
            ]
                \node[draw,rounded corners=9pt,rectangle,minimum width=17mm, minimum height=17mm] (device) at (0,0) {IoT};
                \node (lupe) at (device.-30) {\faSearch};
                \node[anchor=south] (iot-label) at ($(device.north)+(0,0.4)$) {IoT Device};

                \node[font=\normalsize] (specs) at ($(device)+(4.5,0)$) {\faFile*[regular]};
                \node[anchor=south,align=center] at (specs.north) (spec-label) {Device\\Model};
                \draw[-,very thick] (specs) -- (spec-label);
                
                \draw[myarrow] (device) -- (specs);

                \node[font=\huge] (filter) at ($(specs)+(4,0)$) {\faFilter};
                \draw[myarrow] (specs) -- (filter);

                \node[font=\normalsize] (poll) at ($(filter)+(0,3.5)$) {\faFile*[regular]};
                \node[anchor=south] at (poll.north) {Testing Profile};
                \draw[myarrow] (poll) -- (filter);

                \node[font=\normalsize,align=center] (catalogue) at ($(filter)+(4,0)$) {\faFile*[regular]\,\faFile*[regular]\\\;\faFile*[regular]\,\faFile*[regular]};
                \node[anchor=south,align=center] at (catalogue.north) {Test Case\\Catalog};
                \draw[myarrow] (catalogue) -- (filter);

                \node[font=\normalsize] (exe) at ($(filter)+(0,-3)$) {\faCogs};
                \node[anchor=west,align=center] at (exe.east) {Test Protocol\\Execution};
                \draw[myarrow] (filter) -- (exe);

                \node[font=\normalsize] (result) at ($(exe)+(0,-3)$) {\faFile*[regular]};
                \node[anchor=west] at (result.east) {Test Result};
                \draw[myarrow] (exe) -- (result);

                \node[font=\normalsize, align=center] (stamp) at ($(result)+(0,-2.75)$) {\faStamp\\Assessment};

                \node[text=OliveGreen] (pass) at ($(result)+(2.5,-2.5)$) {\faCheck};
                \node[text=red] (fail) at ($(result)+(-2.5,-2.5)$) {\faTimes};
                \draw[myarrow] (result) -- (pass);
                \draw[myarrow] (result) -- (fail);

                \node[anchor=south west,font=\normalsize] (legend) at (iot-label.west|-stamp) {\faFile*[regular]\;\dots};
                \node[anchor=west,align=center,outer sep=0,inner sep=0] at (legend.east) {machine\\readable};

            \end{tikzpicture}
    \caption{Visualization of the security assessment process}
    \label{fig:certification-process}
\end{figure}

Our proposed security assessment process is inspired by the model-based testing (MBT) approach. MBT refers to the automatable derivation of test cases based on a model of the system or device under test (SUT/DUT)~\cite{utting2012taxonomy} and has been explored in the context of IoT~\cite{lonetti2023model}.
While it would be overly optimistic to expect security assessment to be fully automated, the goal should be to automate as much of the process as possible. Figure~\ref{fig:certification-process} shows the required components of the testing process and their interrelation. In the following, we will briefly present the security assessment process before we examine each step of the process regarding its inputs, outputs, and the opportunities for automation.

The foundation of an automated security assessment process is a comprehensive \textbf{Test Case Catalog} where each case is carefully defined to ensure objectivity and replicability. 
Due to the heterogeneity of IoT devices, not all tests are suitable for every IoT device, and therefore, a \textbf{Filter Process} is required to select the applicable test cases.
Therefore, it is necessary to derive a sufficiently accurate abstraction of the device (\textbf{Device Model}) that allows for a systematic testing approach. 
Further, a \textbf{Testing Profile} formalizes the assumptions under which a device is tested. For example, the OWASP IoT Security Testing Guideline (ISTG)~\cite{owasp-ISTG} defines physical access and authorization levels to which specific test cases map. Introducing further categories to cover, e.g., data sensitivity or security impact of different IoT devices, would further make the selected test cases more suitable. 
The purpose becomes obvious when comparing a smart bulb with a smart door lock. While the malfunction of a bulb leads to inconvenience for the user, the malfunction of the lock might lead to severe consequences. Similarly, the data generated by the bulb is of minor sensitivity, while the door lock data might allow the identification of patterns that malicious actors could abuse.

An array of test cases is selected based on the device model and testing profile. This requires each test case to have properties assigned for each testing profile and device model characteristic.
The next step is the \textbf{Execution} of the selected tests according to the defined protocols followed by the \textbf{Assessment} of the results which leads to the overall security assessment of the device.
This leaves us with three  (filtering, test case execution, and test result assessment) and two auxiliary (creation of the device model and testing profile) processes.

The \textbf{Filter Process} takes as its \textit{inputs} the device model, the testing profile, and the test case catalog. Mapping the relevant device characteristics (e.g., single binary blob versus embedded Linux, or WiFi versus Bluetooth) with the specified testing criteria (e.g., light bulb versus smart lock), reduces the overall catalog down to the pertinent test scenarios. The \textit{output} of this process is a list of test cases, along with step-by-step execution guides that reflect the specific context of the DUT. Given that the inputs are available in machine-readable form, this process can be fully automated.

In the \textbf{Test Protocol Execution} step, the \textit{inputs} comprise the previously created list of test cases and the actual DUT. In an iterative process these test cases are executed as per the documented guidance. Although full automation of this stage is impractical, certain test cases might be partially or even fully automated, as has already been explored~\cite{kaksonen2024automating}.
The output of this phase is a list of test execution protocols for each case, which, in turn, should be provided in machine-readable syntax to facilitate automation in the upcoming step.

Finally, the \textbf{Assessment} process uses the test execution protocol as its primary \textit{input} to determine a binary evaluation (pass/fail) for each test case. A structured assessment scheme ensures consistency: certain failures may trigger an automatic fail of the overall security assessment, while other issues might be acceptable within defined thresholds. The \textit{output} of this process is twofold: each test case receives a pass/fail verdict, and the device itself is either assessed as secure or insecure based on the overall results. This process might be fully automated once the assessment scheme is defined comprehensively.

\subsection{The Roadmap}

To facilitate scalable security assessment of IoT devices, we identify the major building blocks that should be addressed in future research.
\subsubsection{Device (Meta) Model}
The MBT approach requires a model of the DUT that is sufficiently detailed while being abstract enough to fit all considered devices. The extent of the model is closely related to the questions of the IoT’s Certification Quagmire, specifically to the scope of the security assessment. Notable work in this direction is the device reference architecture proposed in~\cite{guth2016comparison} and its subsequent development by the OWASP ISTG project~\cite{owasp-ISTG}. While other device models typically are either too abstract or include way too little detail, this model's level of detail seems appropriate for the MBT approach. However, it might be too narrow in its scope, leaving out components essential for a potentially trustworthy security assessment. 

While the ISTG model specifies physical access levels, i.e., the degree of how close attackers can get to the device (ranging from remote access to invasive access), and authorization access levels, i.e., the assumed digital privileges of attackers (ranging from unauthorized access to manufacturer-level access), it fails to include other important dimensions to fully define the potential security implications of the DUT.

Essential dimensions missing from OWASP's device model are a \textit{data security} dimension to capture different levels of data sensitivity and a \textit{security impact} dimension to capture the impact of a security breach of the DUT on the whole system.

Furthermore, manually creating a device model for each specific DUT, as it is done in most approaches so far, e.g.~\cite{kaksonen2024automating}, is for itself a very cumbersome, error-prone, and time-consuming task. This could be automated, at least to a very large extent, by creating and using some kind of meta model that has to be general enough to capture all relevant properties of all possible DUTs and still specific enough such that it is instantiable for a given device with adequate effort. One possible approach to generating an adequate device meta model would be to use an ontology.

\subsubsection{Testing Profile}
Usually, IoT devices are not standalone devices operating solely autonomously. Rather, they are embedded into a whole ecosystem, commonly consisting of a cloud connection, which allows data to be combined and stored permanently, or a mobile app, which allows for interaction with the device and the data stored on it. A testing profile has to identify all relevant backend and control systems with which the DUT usually interacts. Furthermore, the testing profile has to specify some kind of \textit{verification level} that defines the rigor with which the tests are executed, ranging from formal verification to overall verification. This verification level can either be applied to the whole DUT, differ for different components of the DUT, or even be different for single test cases, depending on the requirements, the complexity, and the intended usage of the DUT.

\subsubsection{Test Case Catalog}

Although there exist lists of common IoT vulnerabilities~\cite{owasptopten}, vulnerability databases specifically for IoT devices~\cite{VARIoT_db}, and descriptions of baseline security recommendations for IoT products~\cite{ENISA2017baseline, owasp-ISVS}, sometimes augmented with specific tests~\cite{ETSI2021}, at the present, there is no test case catalog that fully covers the broad spectrum of CIoT devices. Given the heterogeneous nature of IoT devices and that most probably in the future, there will be completely new devices participating in the IoT, such a test case catalog will never be finished or complete but probably ever-growing or evolving. Nonetheless, starting such a catalog that should combine existing work and break requirements down on a fine granular test case level would be a strict prerequisite for large-scale automated security assessment of IoT devices.

Furthermore, for seamless interaction of the test cases with the device model and verification profile, each test case has to include the following prerequisites specifying 1) the physical and authorization access level needed for executing this test case, 2) the data security and security impact level, and 3) for which verification levels the test case has to be included in the list of test cases to be executed.

One possible approach to realize such a test case catalog is to make use of a taxonomy. Although there already exist a number of proposals for IoT taxonomies, e.g.,~\cite{peccs14,IoTUK}, where several dimensions such as \encom{technical complexity}, \encom{safety, security, privacy}, or \encom{data sharing} are exemplified and even such taxonomies specifically for the domain of IoT security exist~\cite{alqassem2014taxonomy,nawir2016internet,Xenofontos.2021}, none of these taxonomies has a strict focus on the testing of security properties of whole classes of IoT devices in a similar manner. 

\subsubsection{Machine Readability}
To achieve the highest possible degree of automation, it is important that everything within the security assessment process, i.e., device model, testing profile, test case catalog, test protocols, test results, and result assessment has to be machine-readable. This has to be present when choosing and applying ontologies for the device (meta) model or when creating a taxonomy for the test case catalog.

\subsubsection{Automation of Test Execution}
Recent work on the automation of IoT security testing~\cite{kaksonen2024automating}, given a manually created device profile and manually selecting the relevant test cases for network-based threats from~\cite{ETSI2021}, shows that basic network security tools alone can automate approximately 52\% of these tests, while advanced tools can raise coverage to around 70\%. Although these numbers might be overly optimistic for all the security assessments of a general IoT device, including its ecosystem, they show the potential of automation.

\section{Conclusion}
\label{sec:conclusion}

Summarizing the above, it is clearly visible that besides the efforts in the scientific community to develop IoT-specific methods for fulfilling the basic protection goals of confidentiality, integrity, and availability on resource-constrained devices, in practice, users of IoT devices usually face a completely different set of risks. This is probably best highlighted with the example that the Top 1 vulnerability listed in OWASP's Top 10 IoT list~\cite{owasptopten} are weak, guessable, or even hardcoded passwords, while researcher now demand \encom{lightweight quantum security schemes} to harden IoT devices against attacks executed by quantum computers~\cite{Bouguettaya2021}. Although this and similar research directions are important and valid for a potential future, no adversary has the incentive to invest in a quantum computer when she can simply guess the password of an IoT device. 

Furthermore, manufacturers currently have hardly any incentive to invest in more secure IoT devices. Despite the widespread use of IoT devices, consumers so far have little awareness of their security and privacy properties and are rarely willing to spend more money on more secure products. Even if this awareness is present, the average user has no clue on which product might be more secure than others. In addition to that, the situations where users can \encom{opt out of the IoT} become increasingly rare~\cite{schneier2017iot}, and even if users can do that for themselves, they are still part of the IoT through other users' IoT devices or concepts like the Smart City. 

One possibility to change that is governmental intervention and more and more governments do exactly that, like the European Commission’s Cyber Resilience Act or the US government's IoT Cybersecurity Improvement Act of 2019~\cite{US-IOTCIA}.  
Some researchers hope that these initiatives help the market to create more secure IoT devices as governmental intervention did for cars, food, airplanes, and many other markets~\cite{schneier2017iot}.

Furthermore, if users can easily assess the security and privacy properties of an IoT device, they might be willing to pay more for more secure IoT products~\cite{emami2023consumers}.

So, we still face the problem of a systematic, reliable, and comparable security assessment of IoT devices without which the compliance of the manufacturers of IoT devices to the governmental guidelines or potential certificates becomes a very cumbersome task. Thus, we have to build an automated and transparent way of systematically assessing the security of consumer IoT devices, as we propose in this paper. Together, this might finally incentivize IoT manufacturers to build in security from the very start.

Keeping in mind the increasing amount of IoT devices already on the market, the time is now to develop and operate a testing and assessment procedure which allows to automatically evaluate the security of Consumer IoT devices. Hereby, we have to keep in mind the heterogeneous nature of IoT devices, including the challenges that arise during testing them~\cite{voas2018testing,Medhat2019} and also the challenges of a following certification of IoT devices~\cite{Voas2018}. Only if we start now, taking into account all the collections of common IoT security vulnerabilities~\cite{owasptopten,ENISA2017baseline,ETSI2021}, we can succeed to make the Internet of Things, Smart Homes, and Smart Cities more secure.

\bibliographystyle{IEEEtran}  
\bibliography{literatur}     

\end{document}